%% file: main.tex
\documentclass[sigconf]{acmart}

\AtBeginDocument{%
  \providecommand\BibTeX{{%
    \normalfont B\kern-0.5em{\scshape i\kern-0.25em b}\kern-0.8em\TeX}}}
\setcopyright{acmcopyright}
\copyrightyear{2023}
\acmYear{2023}
\acmDOI{XXXXXXX.XXXXXXX}

\acmConference[ICSE 2024]{46th International Conference on Software Engineering}{April 2024}{Lisbon, Portugal}

\usepackage{hyperref}      
\usepackage{url}            
\usepackage{booktabs}       
\usepackage{amsfonts}       
\usepackage{nicefrac}       
\usepackage{microtype}      
\usepackage{lipsum}		    
\usepackage{doi}
\usepackage{subcaption}
\usepackage{colortbl}
\usepackage{balance}
\usepackage{xspace}
\usepackage{acronym}
\usepackage{adjustbox}
\usepackage{siunitx}
\usepackage{tabularx}
\usepackage{enumitem}

\newcommand{\az}[0]{\textsc{AndroZoo}\xspace}
\newcommand{\bestAriScoreDescription}[0]{\num{0.91}\xspace}
\newcommand{\datasetName}[0]{\textsc{AndroCatSet}\xspace}
\newcommand{\approachName}[0]{\textsc{G-CatA}\xspace}

\usepackage[strict]{changepage}
\usepackage{framed}
\definecolor{formalshade}{rgb}{0.85,1,0.85}
\definecolor{darkblue}{rgb}{0.0,0.6,0.30}
\newenvironment{formal}{%
  \MakeFramed{\advance\hsize-\width\FrameRestore}%
  \noindent\hspace{-4.55pt}
  \begin{adjustwidth}{}{7pt}%
}
{%
  \end{adjustwidth}\endMakeFramed%
}

\newboolean{showcomments}
\setboolean{showcomments}{true}
\ifthenelse{\boolean{showcomments}}
 { \newcommand{\mynote}[2]{
      \fbox{\bfseries\sffamily\scriptsize#1}
        {\small$\blacktriangleright$\textsf{\emph{#2}}$\blacktriangleleft$}}}
        { \newcommand{\mynote}[2]{}}

\begin{document}
\title[Revisiting Android App Categorization]{Revisiting Android App Categorization}

\author{Marco Alecci}
\affiliation{%
  \institution{SnT, University of Luxembourg, \\ Luxembourg, Luxembourg}
  \country{}
}
\email{marco.alecci@uni.lu}

\author{Jordan Samhi}
\affiliation{%
  \institution{CISPA Helmholtz Center for Information Security \\ Saarbrücken, Germany}
  \country{}
}
\email{jordan.samhi@cispa.de}

\author{Tegawendé F. Bissyandé}
\affiliation{%
  \institution{SnT, University of Luxembourg, \\ Luxembourg, Luxembourg}
  \country{}
}
\email{tegawende.bissyande@uni.lu}

\author{Jacques Klein}
\affiliation{%
  \institution{SnT, University of Luxembourg, \\ Luxembourg, Luxembourg}
  \country{}
}
\email{jacques.klein@uni.lu}

\begin{abstract}
\input{Sections/00_Abstract.tex}

\end{abstract}



\maketitle

\input{Sections/01_Introduction.tex}

\input{Sections/02_Background}
\input{Sections/03_DatasetCreation}
\input{Sections/04_ExistingApproaches}

\input{Sections/05_ImprovingDescriptionBased}

\input{Sections/06_ImprovingApkBased}
\input{Sections/07_DownstreamTasks}

\input{Sections/08_Limitations}
\input{Sections/09_RelatedWork}
\input{Sections/10_Conclusion}
\input{Sections/11_DataAvailability}
\input{Sections/12_Acknowledgement}

\balance
\bibliographystyle{ACM-Reference-Format}
\bibliography{references}

\end{document}

%% file: Sections/00_Abstract.tex
Numerous tools rely on automatic categorization of Android apps as part of their methodology. However, incorrect categorization can lead to inaccurate outcomes, such as a malware detector wrongly flagging a benign app as malicious. One such example is the \textit{SlideIT Free Keyboard} app, which has over \num{500000} downloads on Google Play. Despite being a "Keyboard" app, it is often wrongly categorized alongside "Language" apps due to the app's description focusing heavily on language support, resulting in incorrect analysis outcomes, including mislabeling it as a potential malware when it is actually a benign app.
Hence, there is a need to improve the categorization of Android apps to benefit all the tools relying on it.

In this paper, we present a comprehensive evaluation of existing Android app categorization approaches using our new ground-truth dataset. Our evaluation demonstrates the notable superiority of approaches that utilize app descriptions over those solely relying on data extracted from the APK file, while also leaving space for potential improvement in the former category. Thus, we propose two innovative approaches that effectively outperform the performance of existing methods in both description-based and APK-based methodologies. Finally, by employing our novel description-based approach, we have successfully demonstrated that adopting a higher-performing categorization method can significantly benefit tools reliant on app categorization, leading to an improvement in their overall performance. This highlights the significance of developing advanced and efficient app categorization methodologies for improved results in software engineering tasks.

%% file: Sections/01_Introduction.tex
\section{Introduction}
\label{sec:introduction}

Automatic categorization of Android apps has proven to be a useful tool for addressing numerous software engineering tasks. The applications of categorization encompass a wide range of tasks, including but not limited to anomaly detection~\cite{gorla-chabadaicse-2014, avdiienko2017, zhang2018}, detection of miscategorized apps~\cite{aminordin2018, surian2017}, malware detection~\cite{YANG201727,mahindru2021}, or classifying malware into distinct families~\cite{li2022, narayanan2018, nix2017}.

One of the most popular examples is the CHABADA framework developed by Gorla et al.~\cite{gorla-chabadaicse-2014}, which focuses on detecting applications that exhibit behavior inconsistent with their provided descriptions in order to find potential malicious apps. CHABADA leverages app descriptions to categorize Android apps, and subsequently employs unsupervised One-Class SVM anomaly detection to identify outliers based on API usage patterns. However, an inadequate categorization approach can significantly impact the overall accuracy of CHABADA, leading to numerous false positives (i.e., a goodware detected as a malware) and/or false negatives. For example, the authors themselves acknowledged in their paper that the app \textit{SlideIT Free Keyboard}, which allows users to insert text by sliding their finger along the keyboard, was categorized alongside language apps because over half of the app's description focuses on language support. The app was wrongly detected as an anomaly and potential malware, despite its harmless nature, due to the wrong categorization. However, without a manual inspection, similar to the one conducted by the authors of CHABADA for this app, it becomes challenging to determine whether factors like the number of false positives are primarily influenced by the app categorization module or the anomaly detection phase, as we can only evaluate the combined effect of both.

Inaccurate categorization, as seen in \textit{SlideIT Free Keyboard}, underscores the need for a more precise categorization to minimize such errors. Hence, as a starting point, we initiated our research on automatic app categorization with a comprehensive literature search to gather existing categorization approaches with the primary aim of evaluating and effectively comparing their performance. However, during our attempt to compare the existing approaches that we retrieved, we encountered a significant obstacle: \textit{how can we evaluate and compare their performances?} Indeed, despite the extensive research conducted on automatic app categorization, there remains a lack of a  ground-truth dataset, as highlighted in previous studies~\cite{alsubaihin2019, ebrahimi2021}. Thus, this hinders the evaluation and comparison of different categorization approaches.

Some researchers attempted to overcome this limitation by using Google Play categories as a basis for evaluating their approaches ~\cite{li2022, olabenjo2016applying}. However, both previous studies and our own research (Section ~\ref{subsec:rq0}), demonstrate that apps within the same Google Play category exhibit only a broad sense of similarity, with significant variations in granularity observed across different categories. ~\cite{alsubaihin2016, martin2017, surian2017, gorla-chabadaicse-2014}. For example, the \texttt{HOUSE\_AND\_HOME} category on Google Play encompasses apps for buying/renting houses and controlling Smart Home IoT devices. Although both relate to "House," their purposes differ, rendering this categorization unsuitable as a ground-truth. 

Therefore, a reasonable and up-to-date alternative to evaluating a categorization approach is to create a dataset that relies on human judgment to manually categorize the apps. Previous researchers have evaluated their approaches on datasets that were manually crafted but often limited in either the number of apps, like Subaihin et al. ~\cite{alsubaihin2019} which only manually analyzed 300 apps in their research, or in the number of categories, with Ebrahimi et al. ~\cite{ebrahimi2021} testing only two categories. In contrast, our paper addresses this limitation by developing \datasetName: the first ground-truth dataset for app categorization, which comprises \num{5000} apps categorized into 50 classes. To prevent confusion with Google Play's "categories," we use the term "classes" to define clusters of apps with shared purposes and functionalities (e.g., calculators, navigation apps, weather apps, etc.). 

Leveraging our new ground truth \datasetName, we were able to confirm, at first, the observations made in prior studies concerning the inefficiency of the Google Play Store categorization schema~\cite{alsubaihin2016, martin2017, surian2017, gorla-chabadaicse-2014} and secondarily, compare the effectiveness of current categorization approaches. Our comparison of existing approaches has revealed the remarkable superiority of categorization methods that leverage app descriptions compared to those that rely solely on data extracted from the APK file, such as method names or strings used by the app. However, despite the success of description-based approaches, our evaluation also revealed that there is still room for improvement within these methods. Therefore, we developed two innovative approaches: one description-based and the other APK-based, and thoroughly evaluated them using \datasetName, providing a comparison with existing methodologies. The results of our evaluation demonstrated a significant improvement in performance for both the description-based and APK-based methodologies, as compared to the existing approaches. 

One particularly noteworthy approach is our new description-based method, which we named \approachName: \textit{\textbf{G}pt-based \textbf{CAT}egorization of \textbf{A}ndroid apps}, as it relies on Gpt-based models developed by OpenAI\footnote{https://openai.com/} (as we will describe in detail in Section~\ref{subsec:rq2}). When using \approachName, the app \textit{SlideIT Free Keyboard}, previously discussed, can be appropriately categorized alongside other keyboard apps. This correction addresses the error made by CHABADA, which erroneously categorized it under language apps, thus highlighting how \approachName can benefit tools relying on automatic categorization like CHABADA. To prove this point further, we conducted an additional experiment where we implemented the complete CHABADA framework, incorporating both the original and \approachName approaches to categorize apps. Results indicate that our approach \approachName exhibited greater precision in detecting anomalies compared to the original CHABADA approach.

The main contributions of our work are as follows:
\begin{itemize}[noitemsep,topsep=0pt]
    \item We release \datasetName: the first ground-truth dataset for evaluating Android app categorization approaches.
    \item We show that apps belonging to the same Google Play category exhibit only a general sense of similarity, confirming previous studies on the subject.
    \item We conduct a comparative analysis of existing categorization approaches, revealing that description-based approaches outperform APK-based ones.
    \item We propose two novel approaches, one description-based (\approachName) and one APK-based, that improve the performance of existing methods.
    \item We demonstrate that our new description-based approach \approachName can offer significant benefits to tools that depend on app categorization, such as CHABADA.
\end{itemize}
\textbf{Data Availability.} Our ground-truth dataset \datasetName, our new approach \approachName, and all our artifacts are available at: 
\begin{center}
     \href{https://github.com/Trustworthy-Software/Revisiting-Android-App-Categorization}{https:/github.com/Trustworthy-Software/Revisiting-Android-App-Categorization}
\end{center}

%% file: Sections/02_Background.tex
\section{Background}
\label{sec:background}

In this section, we introduce terms and concepts used throughout our paper. First, we describe the clustering evaluation metric we relied on. Second, we provide an overview of text-embedding models, as they are employed in both existing categorization approaches and the new ones we are proposing.

\noindent \textbf{Clustering Evaluation Metrics.}
Clustering evaluation metrics serve to assess the performance of clustering algorithms, and they can be categorized into two primary groups: intrinsic and extrinsic metrics~\cite{Amigó_Gonzalo_Artiles_Verdejo_2009}. 
Intrinsic metrics assess cluster quality and cohesion, focusing on the internal structure of the clusters, without relying on external information or ground truth. Examples of intrinsic metrics include the Silhouette Coefficient~\cite{ROUSSEEUW198753}, Davies-Bouldin Index~\cite{daviesBouldin}, and Calinski-Harabasz Index~\cite{calisnki}. 
On the other side, extrinsic metrics are used to measure the similarity between two data clusterings, where one of the clusterings may consist of a ground-truth made of known labels. Examples of extrinsic metrics include the Rand Index ~\cite{randIndex} and Fowlkes-Mallows Index ~\cite{fowlkesMallows}.

In our paper, we primarily utilize the Adjusted Rand Index (ARI)~\cite{Hubert_Arabie_1985} as our main metric. The ARI is an extrinsic metric that adjusts the Rand Index for chance and yields a value between -1 and 1. A value of 1 indicates perfect agreement between the two clusterings, 0 indicates random agreement, and -1 indicates complete dissimilarity between the two clusterings.
More in detail, given a set of n elements, and two partitions (e.g., clusterings) of these elements, namely $X=\left\{X_1, X_2, \ldots, X_r\right\}$ and $Y=\left\{Y_1, Y_2, \ldots, Y_s\right\}$
the overlap between X and Y can be summarized in a contingency table $\left[n_{ij}\right]$, as shown in (\ref{eq:contingencyTable}). Each entry $n_{ij}$ denotes the number of objects in common between $X_{i}$ and $Y_j$:$n_{i j}=\left|X_i \cap Y_j\right|$.
\begin{equation}
    \begin{array}{c|cccc|c}
    X^Y & Y_1 & Y_2 & \cdots & Y_s & \text { sums } \\
    \hline X_1 & n_{11} & n_{12} & \cdots & n_{1 s} & a_1 \\
    X_2 & n_{21} & n_{22} & \cdots & n_{2 s} & a_2 \\
    \vdots & \vdots & \vdots & \ddots & \vdots & \vdots \\
    X_r & n_{r 1} & n_{r 2} & \cdots & n_{r s} & a_r \\
    \hline \text { sums } & b_1 & b_2 & \cdots & b_s &
    \end{array}
\label{eq:contingencyTable}
\end{equation}
Once the contingency table is defined, the ARI is computed as follows:
\begin{equation}
        A R I (X,Y) = \frac{{\sum_{ij} \binom{{n_{ij}}}{{2}} - \left[\sum_i \binom{{a_i}}{{2}} \sum_j \binom{{b_j}}{{2}}\right] / \binom{{n}}{{2}}}}{{\frac{1}{2} \left[\sum_i \binom{{a_i}}{{2}} + \sum_j \binom{{b_j}}{{2}}\right] - \left[\sum_i \binom{{a_i}}{{2}} \sum_j \binom{{b_j}}{{2}}\right] / \binom{{n}}{{2}}}}
        \label{eq:ari}
\end{equation}
where $n_{ij}$, $a_i$, $b_j$ are values from the contingency table. For instance, in our paper, we consider $X$ as the categorization approach that we want to evaluate and $Y$ as our ground-truth against which we compare the approach.

\noindent \textbf{Text Embedding.}
Text embedding models are an advanced approach to convert textual information into numerical representations. These models, such as Word2Vec ~\cite{word2vec}, FastText~\cite{bojanowski-etal-2017-enriching}, GloVe ~\cite{glove}, and BERT ~\cite{bert}, capture the semantic and syntactic meaning of words, sentences, or entire documents by mapping them to dense vector spaces. While Word2Vec, FastText, and GloVe primarily focus on word-level embeddings, transformer-based models such as BERT can handle not only individual words but also whole sentences or documents. 
Moreover, OpenAI, the company behind ChatGPT, has released their own text embedding models~\cite{embeddingOpenAIblog}, which are readily accessible through their official API. They can be leveraged for numerous applications, including search, clustering, recommendations, anomaly detection, and classification, as stated in the official documentation~\cite{embeddingsOpenAIdocumentation}. The text embedding models are based on pre-trained GPT (Generative Pre-trained Transformer) models~\cite{embeddingOpenAIpaper}, which are a class of transformer-based models developed by OpenAI~\cite{gptPaper}.
These models, such as GPT-3 or the most recent GPT-4 \cite{gpt4Paper}, are built upon the powerful transformer architecture and have achieved significant breakthroughs across various NLP tasks~\cite{gptPaper}.

%% file: Sections/03_DatasetCreation.tex
\section{Ground-Truth Dataset}
Existing studies have brought attention to the lack of a  ground-truth dataset for assessing categorization approaches~\cite{alsubaihin2019, ebrahimi2021}. Furthermore, additional studies have pointed out that utilizing Google Play categories for such evaluations is unsuitable since apps within the same Google Play category demonstrate only a broad sense of similarity, with significant variations in granularity across the different categories ~\cite{alsubaihin2016, martin2017, surian2017, gorla-chabadaicse-2014}.

To bridge this gap, we developed \datasetName: the first ground-truth dataset, specifically designed to evaluate categorization approaches, comprising \num{5000} apps distributed across 50 classes, with 100 apps for each class. Unlike Google Play's utilization of the term "category," we have opted for the term "class" to denote a cluster of applications that share a common purpose and similar functionalities. Some examples are calculator apps, navigation apps, weather apps, and more. Throughout this paper, we will consistently use the term "classes" to refer to the distinct groups of apps within our ground-truth dataset. This choice is deliberate, aimed at distinguishing them from the Google Play "categories," despite their similar meaning. 

In our dataset, we have collected information for each app, including its package name, the Google Play Category ID (which uniquely identifies the app's category), and the app's description. Table~\ref{table:groundTruthClasses} provides a comprehensive list of all defined classes in the dataset. On average, the apps have an APK size of around 20.38 MB, with a standard deviation of 18.19, indicating variations in size across the apps. The average length of app descriptions is approximately 2083 characters, with a standard deviation of 1087, suggesting differences in the level of detail provided (bearing in mind Google Play's 4000-character limit for descriptions). For more details, a table showing the average APK size and average description length for each class can be found in the repository.

In Section~\ref{subsec:datasetCreation}, we outline, step by step, the process of constructing our new ground-truth dataset. Then, in Section~\ref{subsec:rq0}, we provide some interesting insights about the (non)alignment between app categorization in the Google Play Store and \datasetName.

\begin{table*}[]
\caption{The 50 classes present in our ground-truth dataset.}
\begin{adjustbox}{width=0.8\linewidth,center}
\begin{tabular}{lllll}
\multicolumn{5}{c}{\textbf{Ground-Truth Classes}}                                                          \\ \hline
Airlines                & Alarm                 & Antivirus      & Astrology          & Banking            \\
BarcodeAndQRcodeScanner & BikeAndScooterSharing & BooksReader    & Browser            & BuyAndRentHome     \\
Calculator              & Calendar              & CarBuying      & Dating             & Dialer             \\
Drawing                 & Email                 & ExpenseTracker & FileManager        & FoodDelivery       \\
FoodDiary               & HikingAndTrekking     & Insurance      & Investment         & JobSearch          \\
Keyboard                & LanguageLearning      & Launcher       & Messenger          & MusicProduction    \\
Navigator               & News                  & Notepad        & OnlineTravelAgency & PhotoEditor        \\
PromoAndDeals           & PublicTransit         & Radio          & Recipes            & RemoteControl      \\
Shopping                & SmartHome             & Streaming      & Translator         & TravelGuide        \\
VideoPlayer             & Vpn                   & Wallpaper      & Weather            & WorkoutAndTraining
\end{tabular}
\end{adjustbox}
\label{table:groundTruthClasses}
\end{table*}

\subsection{Dataset Creation}
\label{subsec:datasetCreation}
\begin{figure}
    \centering
    \includegraphics[width=\linewidth]{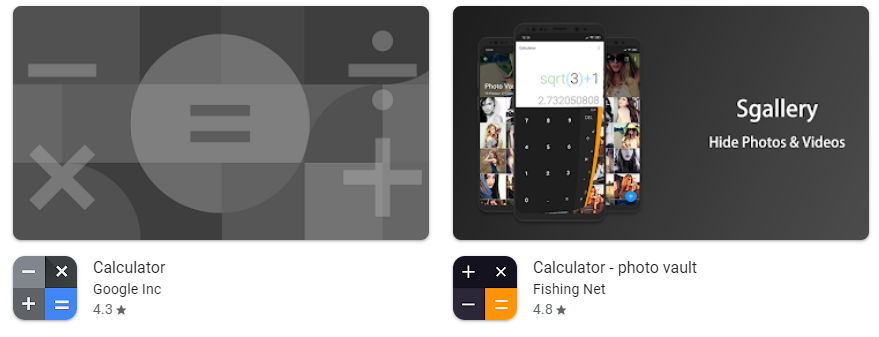}
    \vspace{-6mm}
    \caption{Searching for a calculator app on Google Play often leads to a mix of genuine and fake ones.}
    \label{fig:calculatorsExample}
\end{figure}

Due to the need for high-quality data, relying on a fully automated process for creating the dataset was not feasible. Indeed, simply searching on Google Play for apps with the desired purpose, e.g., searching for "calculator", would have been inadequate due to the presence of apps intentionally disguising themselves as something else. For example, conducting a search for "calculator" on Google Play would yield numerous apps that appear to be regular calculators but actually allow users to hide their photos within them, as illustrated in Figure~\ref{fig:calculatorsExample}. This highlights the importance of implementing a manual verification process during the dataset construction phase, as each app requires individual manual checking.

The dataset was created according to the following steps, which were repeated for each one of the 50 classes: 
\begin{enumerate}[leftmargin=*]
    \item \textbf{Class Definition.}
    We define a class, e.g. "calculator", trying to be as specific as possible to ensure fine granularity with respect to Google Play Categorization. Some examples of classes include weather apps, translator apps, dating apps, and more. We intentionally decided not to define any classes related to games because their functionalities tend to vary too much, requiring excessive manual work to build our ground truth.
    \item \textbf{Apps Collection.}
    We use a custom Python script leveraging the \textit{google-play-scraper}\footnote{https://github.com/JoMingyu/google-play-scraper} to gather apps from Google Play by searching for keywords related to the class we want to populate. As \textit{google-play-scraper} restricts the maximum number of apps returned to 30, it is advisable to search for multiple keywords for each class to gather a more comprehensive set of apps. For instance, in the case of the "Notepad" class, we conduct searches for both "notebook" and "notepad". Additionally, to overcome this limitation, we leverage two parameters of \textit{google-play-scraper}: \texttt{country} and \texttt{lang}. The \texttt{country} parameter represents the two-letter country code (defaulted to "us") used for application retrieval, while the \texttt{lang} parameter denotes the two-letter language code (defaulted to "en") to be used. By modifying the \texttt{country} parameter while keeping \texttt{lang} fixed at "en", we retrieve more applications for each class, ensuring consistency with only English descriptions.
    \item \textbf{Filtering out malicious apps.}
    To ensure that each app in our class functions exactly as intended, we carefully exclude any possible malicious behavior that could alter its functionality and misalign the app with the rest of the class.
    To accomplish this, we search for the package name of each app in the \az dataset~\cite{androzoo2016} and keep only apps with a VirusTotal~\cite{virustotal} score equal to zero, i.e., benign apps.
    \item \textbf{Manual Verification.}
    We then conduct a manual verification process for each app that remains after the previous steps, to determine whether it qualifies as a member of the respective class. This involves a thorough examination of the Google Play page, including the app's name, description, screenshots, and user reviews. In instances of uncertainty, we took an extra step by installing the app ourselves and personally verifying its functionalities. We stop the manual verification process once we reach a total of 100 apps for each class.
    \item \textbf{Download and Information Retrieving.}
     We complete the process by downloading all the apps from \az \cite{androzoo2016}. Moreover, we save in a file the package name, the Google Play Category ID, and the description of the app, previously retrieved using \textit{google-play-scraper}.
\end{enumerate}


\subsection{Dataset Insights.}
\label{subsec:rq0}

After creating the \datasetName ground-truth, we conducted an analysis of the categories assigned by the Google Play Store to the apps included in it. Now, we present the insights yielded by this analysis regarding the granularity level of Google Play's categorization as well as instances of miscategorized apps that were revealed.

\begin{figure*}[t]
    \begin{subfigure}{0.95\linewidth}
        \centering
        \begin{adjustbox}{width=.9\textwidth,center}
        \includegraphics[width=\textwidth]{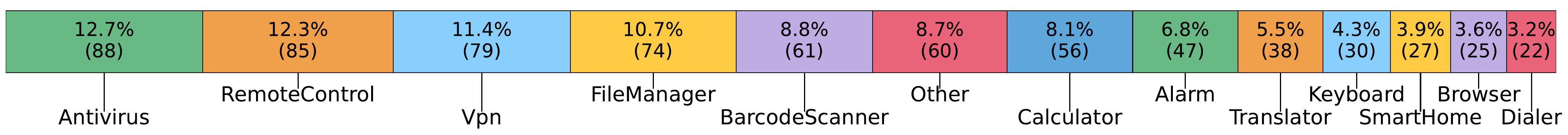}
        \end{adjustbox}
        \caption{\texttt{TOOLS.}}
        \label{subfig:toolsOverview}
    \end{subfigure}
    \begin{subfigure}{0.95\linewidth}
        \centering
        \begin{adjustbox}{width=.9\textwidth,center}
        \includegraphics[width=\textwidth]{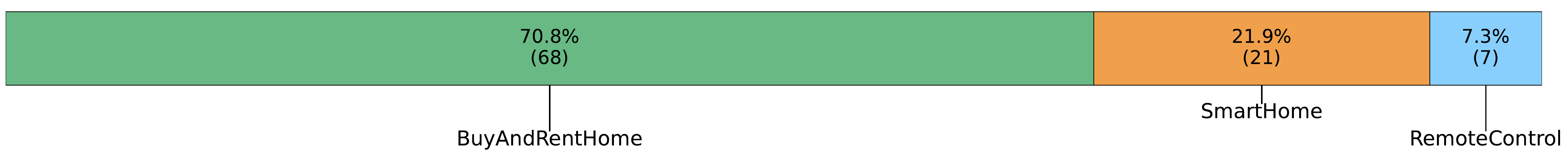}
        \end{adjustbox}
        \caption{\texttt{HOUSE\_AND\_HOME.}}
        \label{subfig:houseAndHomeOverview}
    \end{subfigure}
\caption{Number of apps assigned to a Google Play Store category, segmented into the classes they belong to within our dataset.}
\label{fig:overviewRQ0}
\end{figure*}

\textbf{Granularity Level.}
In Figure~\ref{fig:overviewRQ0}, we focus on the Google Play categories \texttt{TOOLS} and \texttt{HOUSE\_AND\_HOME}. For each category, we provide the count of apps assigned to that specific category, segmented into the classes they belong to within \datasetName. Regarding the \texttt{TOOLS} category (Fig.~\ref{subfig:toolsOverview}), our ground-truth includes a total of 690 apps distributed among 28 distinct classes. To enhance visual clarity, we have grouped classes that constitute less than 3\% of the total under the "Other" class. 

We can clearly see a contrast in the level of granularity between the two Play Store categories, emphasizing that apps within the same category only share a general sense of similarity, as noted in previous studies ~\cite{alsubaihin2016, martin2017}. The Google Play Category \texttt{TOOL} serves as a perfect illustration of this phenomenon, as it contains numerous apps that, despite falling under the umbrella of "tools", exhibit significant variations in their functionality and similarity. Furthermore, even within the seemingly more specific category \texttt{HOUSE\_AND\_HOME} there exist apps with vastly different purposes. For instance, there are apps available for purchasing and renting houses, as well as apps designed for managing Smart Home IoT Devices. This emphasizes the inadequacy of using the categorization system provided by the Google Play Store, especially in anomaly detection approaches. Indeed, in order to get better results, it is preferable to employ custom categorization approaches, such as the ones relying on the description as already observed by Gorla et al.~\cite{gorla-chabadaicse-2014}.

\textbf{Miscategorization.}
Another issue regarding the current categorization system of the Google Play Store is the presence of miscategorized apps, as previously highlighted by Surian et al.~\cite{surian2017}. Upon reviewing the apps within our ground-truth data under the Google Play Category \texttt{TOOLS}, we have identified a clear instance of a miscategorized app in the Google Play Store. The app in question, \textit{Daily Weather}\footnote{https://play.google.com/store/apps/details?id=com.rlk.weathers}, functions as a weather forecast app. In our ground-truth, we appropriately manually assigned it to the "weather" class. However, the app is currently listed under the Google Play Category \texttt{TOOLS}, which indicates a case of miscategorization since the category \texttt{WEATHER} already exists in the Google Play Store. For instance, 98\% of the apps classified as "weather" in \datasetName are appropriately categorized under the Google Play category \texttt{WEATHER}.


%% file: Sections/04_ExistingApproaches.tex
\section{Comparison of Existing Approaches.}

In this section, our objective is to compare various existing categorization approaches and address the following research question:
\begin{center}
    \textbf{RQ1: What is the
performance of the existing categorization
approaches on \datasetName?} 
\end{center}
In Section~\ref{subsec:literatureSearch}, we outline our systematic literature search methodology to gather existing categorization approaches from the literature. Section~\ref{subsec:approachSelection} elaborates on our approach selection criteria, explaining how we identified the approaches to be evaluated. Finally, in Section~\ref{subsec:rq1}, we present the evaluation results to answer RQ1.

\subsection{Systematic Literature Search}
\label{subsec:literatureSearch}
We conducted a systematic literature search to explore the various existing methodologies employed for categorizing Android Apps. Our goal is to systematically retrieve existing approaches in a structured manner to evaluate them on our ground-truth dataset. We conducted our search by exploring three widely used paper repositories: ACM Digital Library~\cite{acmDigitalLibrary}, IEEE Xplore~\cite{ieeeXplore}, and Google Scholar~\cite{googleScholar}.
To ensure comprehensive coverage, we formulated a set of six queries. These queries comprised terms connected by the logical operator "AND," mandating the presence of both terms and terms enclosed in quotation marks, specifying an exact match. The following are the queries we utilized:
\begin{itemize}[leftmargin=*]
    \item android AND "apps categorization"
    \item android AND "apps clustering"
    \item"android app" AND "topic modeling"
    \item"android app" AND "description analysis"
    \item"android app" AND "behavior classification"
    \item"android app" AND "miscategorization"
\end{itemize}

\textbf{Results.}
In Table~\ref{table:literatureSearch}, we present the outcomes of our literature search. The \textit{Hits} column indicates the total number of papers gathered during the process. We then proceeded to eliminate irrelevant publications. \textit{Title} represents the number of papers after removing publications that are clearly irrelevant based on the title. \textit{Abstract} indicates the papers remaining after inspecting the abstracts and removing the not relevant papers. In the final row of the table, we present the overall count of papers corresponding to each step. Finally, we eliminated duplicate papers identified by multiple queries, obtaining a collection of 40 distinct papers.

\begin{table}[h]
    \caption{Paper filtered for each step of the literature search.}
    \begin{adjustbox}{width=.9\columnwidth,center}
    \begin{tabular}{l|ccc}
    \multicolumn{1}{c|}{\textbf{Query}} & \multicolumn{1}{l}{\textbf{Hits}} & \multicolumn{1}{l}{\textbf{Title}} & \multicolumn{1}{l}{\textbf{Abstract}} \\ \hline
    \textit{android   AND "apps categorization"}           & 170 & 46  & 22          \\
    \textit{android AND "app   clustering"}                 & 69  & 26  & 14          \\
    \textit{"android app" AND  "topic modelling"}            & 236 & 45  & 17          \\
    \textit{"android app" AND  "description analysis"}      & 62  & 19  & 12          \\
    \textit{"android app" AND  "behavior classification"}   & 88  & 14  & 3           \\
    \textit{"android app" AND  "miscategorization"}         & 34  & 12  & 2           \\ \hline
    Total                                                   & 659 & 162 & 66         \\
    Removing Duplicates                                     &     &     & \textbf{40}
    \end{tabular}
    \end{adjustbox}
\label{table:literatureSearch}
\end{table}

\textbf{Literature Search Insights.}
In our retrieved collection of papers, we have come across surveys that provide valuable insights into multiple categorization approaches simultaneously~\cite{martin2017, sen2021}. We have discovered numerous papers that utilize app categorization for a variety of purposes, including but not limited to malware detection~\cite{YANG201727,gorla-chabadaicse-2014}, finding miscategorized apps ~\cite{aminordin2018, surian2017}, and even classifying malware into distinct families~\cite{nix2017,li2022,narayanan2018}. As mentioned earlier in Section \ref{sec:introduction}, these findings reinforce the significance and practicality of app categorization. 

Within the collection of papers we have gathered, two major trends can be identified. The first trend centers around the absence of a reliable ground-truth for evaluating app categorization approaches. Numerous papers rely on the Google Play categories~\cite{li2022, olabenjo2016applying, gruschka2018}, but as previous studies have indicated and as we discussed in Section~\ref{subsec:rq0}, these categories are overly generalized to be considered accurate~\cite{martin2017, alsubaihin2016}. Additionally, some papers compare their approaches using the same dataset used by others, which hampers the research community's ability to gain a comprehensive overview based on a consistent dataset. For instance, Shamsujjoha et al.~\cite{shamsujjoha2021} attempted to compare their approach, REACT, against CHABADA~\cite{gorla-chabadaicse-2014}, but they were only able to retrieve only 75\% of the apps utilized in the study by Gorla et al.~\cite{gorla-chabadaicse-2014}. Others have attempted to overcome the lack of a ground-truth by relying on human judgment, but the inherent time constraints associated with this method only allow for relatively small datasets to be manually verified. For example, Ebrahimi et al.~\cite{ebrahimi2021} conducted a study involving 600 apps divided into two categories, whereas Al-Subaihin et al.~\cite{alsubaihin2016} manually evaluated 300 apps in their research. 
The second trend pertains to the reliance of several papers on a supervised approach, often utilizing Google Play categories as labels~\cite{ebrahimi2020, WANG2018987, Fan2019DroidARAAA, rungta2020}. As a result, the number of unsupervised approaches employed for app categorization is significantly reduced.

\subsection{Approaches Selection.}
\label{subsec:approachSelection}

To identify existing categorization approaches to be evaluated on our ground-truth dataset, we began by re-filtering the 40 distinct papers related to app categorization.
Out of the initial 40 papers, only 28 propose a specific approach for app categorization, while the remaining papers consist of reviews discussing multiple approaches or papers relying directly on the categorization already provided by Google Play.
Among the 28 remaining papers, only 9 included available code for implementation. However, the links provided for 3 out of the 9 papers were broken~\cite{narayanan2018, surian2017, chen1025}. Despite making several attempts to contact the authors, we either received no response, or communication ceased after a few emails. Moreover, we tried reaching out to some of the authors of papers that initially did not share their code, but unfortunately, these attempts were also unsuccessful.
For our last step, we excluded one~\cite{kuznetsov2016} of the 6 remaining papers since its authors overlapped with another paper and utilized the same approach (the provided link led to the same web page\cite{appMiningSaarland}).
In the repository, a table showcasing the code availability, utilized features, and reasons for exclusion for each of the 40 papers can be found, providing more in-depth information.

Eventually, the five approaches selected are the following:
\begin{itemize}[leftmargin=*]
    \item \textbf{CHABADA by Gorla et.al~\cite{gorla-chabadaicse-2014}.}
    The objective of CHABADA is to identify applications that do not align with their descriptions. To categorize the apps, they follow a two-step process: clustering the apps using LDA to extract topics from descriptions, and applying K-means for further clustering.
    \item \textbf{REACT by Shamsujjoha et al.~\cite{shamsujjoha2021}.}
    In this paper, the authors present REACT as an alternative approach to CHABADA for scenarios where descriptions are unavailable. Their proposed method follows the same strategy as CHABADA but leverages different sources of information, such as method names, XML data, and GUI text values.
    \item \textbf{Ebrahimi et al.~\cite{ebrahimi2021}.}
    The authors conducted a comparison of various word embedding models to create numerical semantic representations of app descriptions. These representations were then utilized for the classification of app categories. As our paper primarily focuses on the unsupervised categorization of apps, we adopted the initial part of their methodology and applied K-Means clustering on the embeddings to compare it with other approaches. Out of the word embedding models utilized by the authors, we specifically chose GloVe~\cite{glove}, as they demonstrated that it yielded the most favorable outcomes.
    \item \textbf{Sun et al.~\cite{sun2020}}
    The authors introduced a novel mobile app clustering scheme that utilizes various features extracted from the APK file. These features include activity names, certificate issuer information, and sets of keywords. By employing Affinity Propagation~\cite{frey2007}, they demonstrated that clustering based on the similarity of keywords, which represent app functionality, yields superior performance. Hence, for our evaluation, we exclusively focused on the keyword set of strings.
    \item \textbf{Yang et al.~\cite{YANG201727}}
    In this paper, the authors introduce a novel method for characterizing malicious apps by analyzing their descriptions and data-flow information. They employ LDA to extract topics from app descriptions. However, unlike CHABADA~\cite{gorla-chabadaicse-2014}, they directly cluster apps based on the most relevant topic.
\end{itemize}

\subsection{RQ1 Results.}
\label{subsec:rq1}

To assess the performance of the selected approaches, we rely on the ARI Score, which we extensively discussed in Section~\ref{sec:background}. The ARI Score is a straightforward metric that provides a measure of how well the categorization approach aligns with a ground truth. A score closer to 1 indicates a higher overlap with the ground truth and, therefore, better performance, while a score closer to 0 suggests random clustering and poorer performance.

\begin{figure}[h]
    \centering
    \begin{adjustbox}{width=.8\columnwidth,center}
    \includegraphics[width=0.9\linewidth]{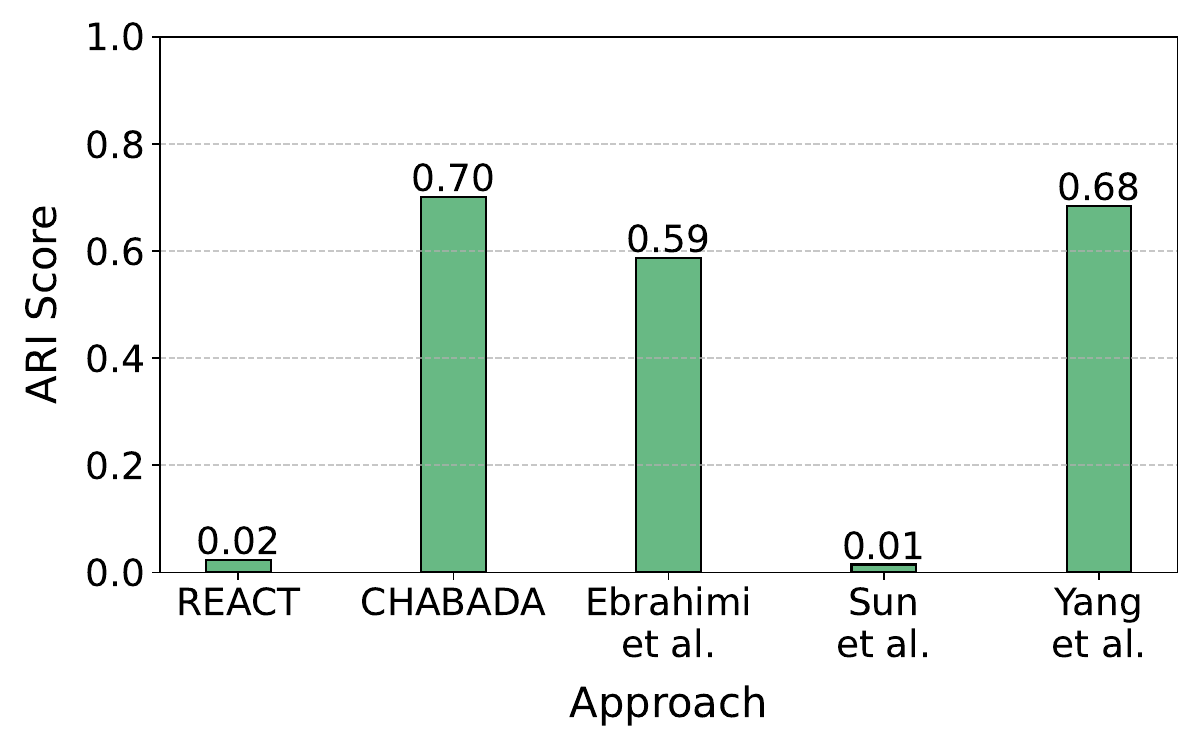}
    \end{adjustbox}
    \caption{ARI Scores of existing categorization approaches.}
    \label{fig:overviewRQ1}
\end{figure}

In Figure~\ref{fig:overviewRQ1}, we present the ARI Score for each of the five approaches, showing a clear and significant distinction among the evaluated methods. The approaches that rely on the description (CHABADA and the ones proposed by Ebrahimi et al. and Yang et al.) demonstrate high performance, while the other two approaches (REACT and the one proposed by Sun et al.) exhibit poor performance when evaluated against on \datasetName. Both of the poorly performing approaches do not rely on the description. Instead, they extract data from the APK file of an app, including XML values, Method Names, GUI Text for REACT, and Strings used by the app for the approach from Sun et al.
An explanation for this behavior has already been proposed by the authors of REACT in their paper~\cite{shamsujjoha2021}. They observed that in Android apps, method names and XML data values can be influenced by data obfuscation and encryption. Moreover, as demonstrated in their paper~\cite{shamsujjoha2021}, it is possible for two distinctly different types of Android apps to have a significant overlap in terms of XML data values.

\begin{formal}
\textbf{Answer to RQ1:} \textit{We evaluated five existing approaches using \datasetName. The approaches relying on app descriptions show relatively high performance, while those extracting data from APK files perform poorly.}
\end{formal}

%% file: Sections/05_ImprovingDescriptionBased.tex
\section{Improvements of Description-Based Categorization Approaches}

RQ1 revealed the superiority of description-based approaches but also acknowledged the potential for further improvement within this methodology. Thus, in this section, we explore possible advancements in description-based approaches.
In Section~\ref{subsec:RQ2experimentalSetup}, we introduce \approachName: our innovative method that leverages app descriptions. Then, in Section~\ref{subsec:rq2}, we evaluate our own approach, addressing the following research question:
\begin{center}
    \textbf{RQ2: 
    Can the performance of app categorization approaches leveraging app descriptions be improved?}
\end{center}

\subsection{Novel Description-based Approach.}
\label{subsec:RQ2experimentalSetup}

We propose our approach \approachName, which stands for \textit{\textbf{G}pt-based \textbf{CAT}egorization of \textbf{A}ndroid apps}. The main idea behind our new description-based approach is to leverage OpenAI's powerful GPT-based text embedding models \cite{embeddingOpenAIblog}, which have been introduced in Section \ref{sec:background}, to effectively process and represent app descriptions, enabling automatic categorization of the apps.

Prior to generating the embeddings, we leverage standard NLP techniques to preprocess the app descriptions, following the approach adopted by similar studies~\cite{gorla-chabadaicse-2014, surian2017, alsubaihin2016}. Our preprocessing steps involve removing non-textual items, eliminating stop-words (e.g., "the," "is," "at," etc.), and performing lemmatization. Lemmatization, unlike stemming, considers the grammatical context and aims to produce meaningful and valid base forms (e.g., "caring" to "care" and not "car" like stemming).

After completing the preprocessing stage, we use the OpenAI second-generation model \texttt{text-embedding-ada-002}~\cite{adav2openAI}. This new embedding model significantly outperforms first-generation models initially introduced by OpenAI in various NLP tasks, such as natural language processing and code tasks, as highlighted in OpenAI's official blog post announcement ~\cite{adav2openAI}. This model employs the \texttt{cl100k\_base} tokenizer, which is the same tokenizer used in ChatGPT3.5 and ChatGPT4~\cite{tokenizersOpenAI}. It allows for a maximum input context length of 8192. Notably, the embeddings generated by this model have 1536 dimensions, and their cost is reduced by 90\% compared to first-generation models ~\cite{adav2openAI}.  

Finally, we use the Scikit-learn~\cite{scikit-learn} implementation of the K-Means~\cite{kmeans} clustering algorithm as the final step to partition the apps into 50 clusters. 


\subsection{RQ2 Results.}
\label{subsec:rq2}
\begin{figure}[h]
    \centering
    \begin{adjustbox}{width=.75\columnwidth,center}
    \includegraphics[width=0.95\linewidth]{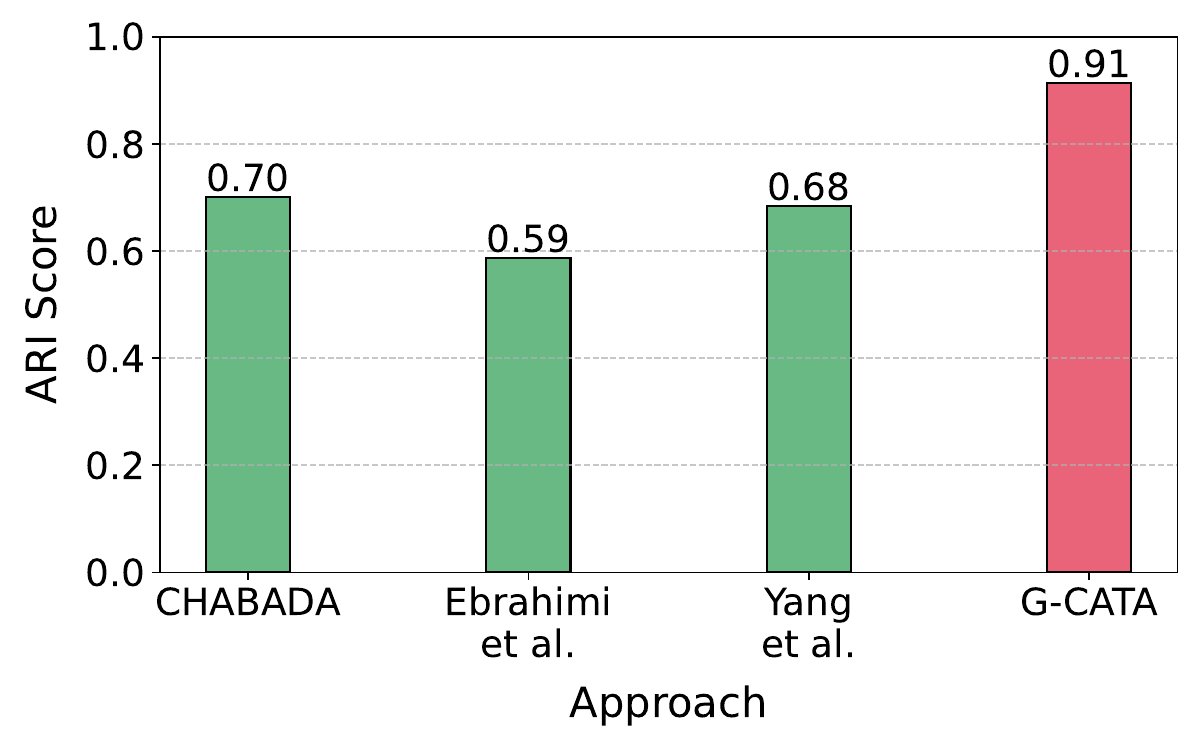}
    \end{adjustbox}
    \caption{ARI scores of our new approach \approachName (red) compared to existing description-based approaches (green).}
    \label{fig:overviewRQ2}
\end{figure}

To address RQ2, we compared our new approach \approachName against existing description-based approaches that were already evaluated in Section~\ref{subsec:rq1}. In Figure ~\ref{fig:overviewRQ2}, we present a comparison of ARI scores among three approaches: \approachName (shown in red) and two existing approaches previously evaluated in RQ1 (shown in green). \approachName outperforms the other two, achieving an impressive ARI score of \bestAriScoreDescription, which represents a remarkable 32\% improvement compared to the previous leading method, CHABADA. Notably, the achieved ARI score is remarkably close to 1, indicating a nearly perfect alignment with the ground truth. This result underscores the effectiveness of combining descriptions with powerful models like OpenAI's text embedding models.

\begin{formal}
\textbf{Answer to RQ2:} \textit{We have proposed \approachName, a novel approach for categorizing Android apps by leveraging app descriptions and OpenAI's powerful embedding models. \approachName achieved an impressive ARI score of \bestAriScoreDescription, outperforming existing description-based approaches.}
\end{formal}


%% file: Sections/06_ImprovingApkBased.tex
\section{Improvements of APK-Based categorization approaches}

RQ1 demonstrated the inadequate performance of categorization approaches relying on APK file data. However, app descriptions may not always be accessible or undergo changes ~\cite{shamsujjoha2021}, emphasizing the need to categorize apps using only data within the APK.
In this section, we explore advancements in APK-based approaches and propose two alternative solutions, both centered around the concept of expanding the set of features used in conventional APK-based methods. These solutions are now presented individually in Section~\ref{subsec:RQ3} and Section~\ref{subsec:RQ4}.

\subsection{Leveraging Existing App Representations from Unrelated Tasks.}
\label{subsec:RQ3}

One simple idea is to leverage existing tools from other unrelated tasks, such as malware detection, to gather diverse app representations (i.e., representations that differ from the one presented in RQ1). These diverse representations can then be used for our categorization task. Let's consider the example of DREBIN~\cite{Arp2014DREBINEA}, one of the most popular Android malware detectors. Instead of using it to distinguish between benign and malicious apps, we can repurpose the same app representation for categorizing apps. This approach requires minimal implementation effort since we can utilize existing features from the malware detection process. 

Thus, we aim to answer the following question:
\begin{center}
    \textbf{RQ3: 
    How do different approaches for app representation, borrowed from unrelated tasks, impact the performance of app categorization?}
\end{center}
To address this RQ, we investigated various app representations proposed in the research literature, aiming to diversify the types of representations explored. In particular, we investigate the following four types of representations that can be generated by available tools.

\begin{itemize}[leftmargin=*]
    \item \textbf{Feature-based app representation.}
    For the "traditional" feature-based representation, we employ DREBIN by Arp et al.~\cite{Arp2014DREBINEA}, which performs extensive static analysis by extracting various application features. These features are then organized in a unified vector space to distinguish between benign and malicious apps. Specifically, DREBIN employs eight distinct feature sets, namely: Hardware features, Requested permissions, App components, Filtered intents, Restricted API calls, Used permissions, Suspicious API calls, and Network addresses.
    \item \textbf{Image-based app representation.}
    The image-based representation is obtained using DexRay by Daoudi et al.~\cite{daoudi2021}. DexRay takes the bytecode from the app's DEX files and transforms them into grayscale images. These images are then utilized by a 1D Convolutional Neural Network (CNN) model to identify malware.
    \item \textbf{Icon-based app representation.}
    To derive this representation, we implement the method proposed by Rajasegaran et al.~\cite{rajasegaran2019}. They introduced a novel icon encoding technique that efficiently detects potential counterfeits for a specific app. This method leverages neural embeddings from CNNs, enhancing search accuracy.
    \item \textbf{BERT-based app representation.}
    To obtain this representation, we adopt LaFiCMIL by Sun et al.~\cite{sun2023laficmil}. The proposed approach, LaFiCMIL, offers a versatile framework that can be effectively applied to diverse BERT-based large file classification tasks, including Android Malware Detection.
\end{itemize}

\textbf{RQ3 Results.}
In Table~\ref{table:overviewRQ3}, we present the ARI scores for the four app representations we previously selected. All four approaches exhibit scores close to zero, which indicates clustering results similar to random chance. Notably, these outcomes align with the two APK-based approaches we discussed in RQ1. Despite our efforts to expand our approaches by leveraging various representations based on diverse features such as app icons, app components, permissions, and bytecode image representations, we were unable to achieve improved performance. However, it is essential to note that we relied on existing tools to extract the app representations from the apps in our dataset, without making any modifications to the feature extraction process or the employed embedding techniques. 

\begin{table}[h]
\caption{ARI scores  using various app representations.}
\vspace{-3mm}
\begin{adjustbox}{width=.9\columnwidth,center}
\begin{tabularx}{\columnwidth}{>{\centering\arraybackslash}X|>{\centering\arraybackslash}X|>{\centering\arraybackslash}X}
\textbf{Approach} & \textbf{Representation} & \textbf{ARI Score} \\
\hline
DREBIN & Features-based & 0.02 \\
DexRay & Image-based & 0.01 \\
Rajasegaran et al. & Icon-based & 0.02 \\
LaFiCMIL & BERT-based & 0.01 \\
\end{tabularx}
\end{adjustbox}
\label{table:overviewRQ3}
\end{table}


\begin{formal}
\textbf{Answer to RQ3:}\textit{
We leveraged various app representations from diverse approaches unrelated to app categorization, e.g., malware detection. Our observations indicate that these representations do not enhance the categorization performance.
}
\end{formal}

\subsection{Novel APK-based Approach.}
\label{subsec:RQ4}
The second option involves creating a new approach from scratch, independent of existing tools, while drawing inspiration from valuable insights offered by the aforementioned app representations used in other, unrelated tasks. In this case, our research question is as follows:
\begin{center}
    \textbf{RQ4: 
    Which specific types of features extracted from the APK file, demonstrate the most significant impact on app categorization performance?}
\end{center}
To answer this question, we first compiled a list of data that can be extracted from an Android APK file and is consistently available, unlike metadata such as the description. Our selected data includes:
\begin{enumerate}[leftmargin=*]
    \item \textbf{Name}: The name/title of the application.
    \item \textbf{Permissions}: The permissions requested by the app.
    \item \textbf{Restricted API Calls}: The API calls protected by permissions.
    \item \textbf{Strings}: The textual content used within the app.
    \item \textbf{Icon}: The graphical representation of the app.
    \item \textbf{Libraries}: The list of libraries utilized by the app.
\end{enumerate}

We have defined this specific list of data with the intention of characterizing apps in a similar manner as the authors of REACT~\cite{shamsujjoha2021} and also taking inspiration from the existing tools that we presented in Section~\ref{subsec:RQ3}. This data is expected to assist us in identifying commonalities among apps with similar purposes. For example, when comparing two "calculator" apps, we expect to find similarities in their app names, rely on a similar set of permissions, and the presence of shared strings within the apps. These shared strings could include terms related to mathematics, such as "calc", "plus", "minus", and similar expressions.

To extract the desired data from the app within our dataset, we have developed a set of custom Python scripts that automate the usage of tools such as Androguard~\cite{androguard} and ApkTool~\cite{apktool}. Our repository also includes these valuable scripts, which are readily available for others to utilize. Androguard, a Python-based software developed for analyzing Android applications, was used to extract essential information, including the app's name, icon, and other pertinent details. In some specific instances, we leveraged ApkTool for reverse engineering Android apps and then parsing XML and DEX files to extract the necessary data.

To handle the diverse nature of the collected information, we employed various techniques to generate numerical feature vectors from the extracted data. When dealing with the names of the apps, we utilized the text embedding models from OpenAI, which were also employed for the descriptions in Section~\ref{subsec:RQ2experimentalSetup}. However, we encountered a limitation with the OpenAI text embedding model's maximum input length, preventing us from generating embeddings for other types of features.
For the app icon, we adopted the same approach of Rajasegaran et al.~\cite{rajasegaran2019}(already presented in Section ~\ref{subsec:RQ3}). 
However, in contrast to their method, we only embedded the content of the app icon, as we believe this provides more value for categorization purposes.
All the remaining data can be conveniently represented as a list of strings. This suggests how a simple TF-IDF vectorization can effectively measure the significance of individual strings, such as permission names or library usage.

Finally, following the approach we employed for the descriptions, we leveraged the K-Means clustering algorithm as the final step to divide the apps into 50 clusters for each of the six types of data we proposed. This allowed us to assess the effectiveness of these features in categorizing the apps on an individual basis.

\begin{figure}[t]
    \centering
    \begin{adjustbox}{width=.85\columnwidth,center}
    \includegraphics[width=\linewidth]{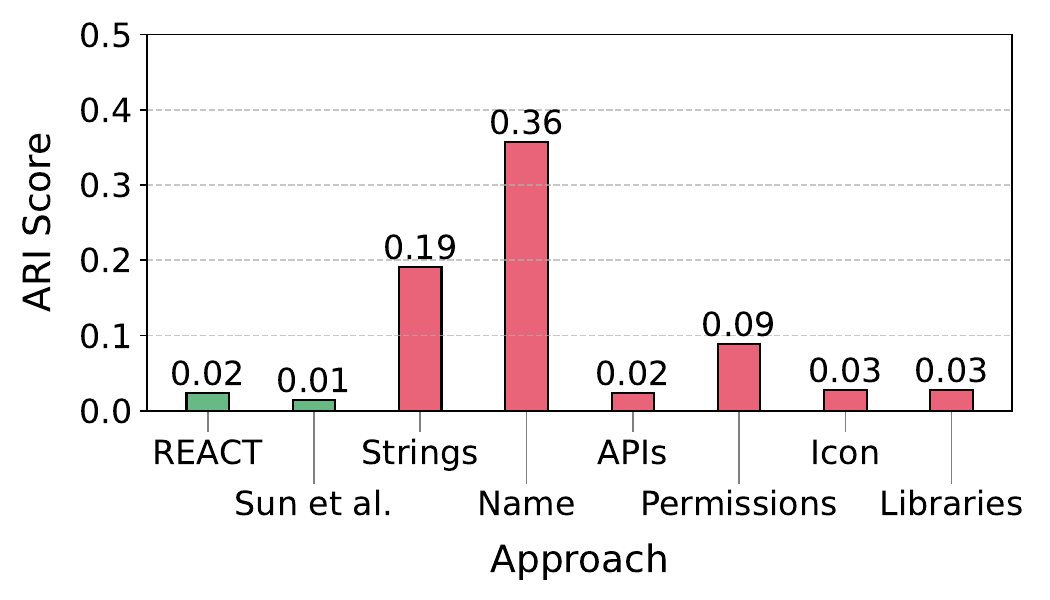}
    \end{adjustbox}
    \vspace{-3mm}
    \caption{ARI scores of our new categorization approach (red) compared to existing APK-based approaches (green).}
    \label{fig:overviewRQ4}
\end{figure}

\textbf{RQ4 Results.}
Figure~\ref{fig:overviewRQ4} illustrates a comparison of ARI Scores for various feature types used in our categorization, represented in red, contrasting with the two existing APK-based methods evaluated in RQ1 shown in green.
As we can observe, using the app icon, the used libraries, and restricted API calls yield results comparable to the two existing APK-based approaches. Instead, notable performance improvements can be achieved when leveraging the app strings, app name, and the requested permissions. Indeed, even if the ARI score obtained through the App Name is still distant from the scores obtained with the description-based approaches, it represents an impressive increase of 1700\% compared to the previous leading APK-based approach, REACT. The remarkable improvement observed can likely be attributed to the utilization of OpenAI's powerful text embedding models, as already discussed for RQ2.

Another key factor worth highlighting is the remarkable performance improvement achieved by leveraging the strings used by the apps, in contrast to the existing APK-based methods that rely on similar characteristics. This emphasizes the substantial influence of embedding techniques and the clustering algorithm on task performance, even when utilizing the same feature set.

\textbf{Combining multiple features extracted from the APK file.}
After evaluating the various types of features individually, as shown in Figure~\ref{fig:overviewRQ4}, and observing their positive impacts, we attempted to further enhance the overall performance by combining all the features simultaneously. 
To achieve this, we concatenated all the previously generated feature vectors and applied normalization using the \texttt{MinMaxScaler} from Scikit-learn~\cite{scikit-learn}. Next, we employed Principal Component Analysis (PCA)~\cite{pca} for dimensionality reduction and subsequently utilized the K-Means algorithm to cluster the apps once more.
Unfortunately, our efforts did not produce the expected improvement. The ARI score obtained was 0.14 when using all six feature types and 0.30 when using only the top three performing ones. These scores did not surpass the best score achieved using only the app name, thus, we decided not to include this score in Figure~\ref{fig:overviewRQ4}. This outcome implies that the different feature types may be capturing similar underlying relationships among the apps. For future research, exploring alternative methods to leverage all features, such as employing ensemble clustering techniques~\cite{BOONGOEN20181}, could be beneficial.

\begin{formal}
\textbf{Answer to RQ4:} \textit{
We tested various APK-based features for app categorization and discovered that combining the app name with OpenAI's model achieved the best performance. Our approach outperformed the leading APK-based method achieving a score of 0.36.
}
\end{formal}

%% file: Sections/07_DownstreamTasks.tex
\section{Downstream Tasks Improvement.}
In Section~\ref{subsec:rq1}, we demonstrated the superiority of our new approach \approachName over existing methods. Building on that, our focus shifts toward examining the potential impact of an enhanced categorization approach on the overall performance of tools reliant on automatic app categorization. We seek to address the following RQ:
\begin{center}
    \textbf{RQ5: How does the performance of tools relying on app categorization improve with a better categorization approach?}
\end{center}
In particular, Section~\ref{subsec:RQ5experimentalSetup}  details the experiments we conducted using \approachName on top of CHABADA.  Then, in Section~\ref{subsec:RQ5}, we provide the findings and answer RQ5.

\subsection{Testing CHABADA as Malware Detector.}
\label{subsec:RQ5experimentalSetup}
CHABADA by Gorla et al.~\cite{gorla-chabadaicse-2014} can be used to find potential malicious apps by identifying inconsistencies between the exhibited behavior and the app descriptions. It consists of two phases: app categorization, which involves utilizing app descriptions, and Latent Dirichlet Allocation (LDA), followed by unsupervised anomaly detection using the One-Class SVM algorithm. This process helps identify outliers based on API usage patterns.

We evaluated the effectiveness of CHABADA in detecting potential malicious apps when employing two different categorization approaches. The objective was to understand how these approaches influence the anomaly detection phase of CHABADA. We trained OC-SVM models on benign apps and used them as classifiers on a test set containing both benign apps and known malware apps, following the methodology of the original paper~\cite{gorla-chabadaicse-2014}. However, we first relied on the original categorization approach based on LDA (i.e., the one presented in [18]) before relaunching all the experiments using our novel approach \approachName.
Below, we provide comprehensive details about the experiments conducted.
\begin{enumerate}[leftmargin=*]
    \item \textbf{Dataset.} We started by organizing \datasetName into a Training Set and a Test Set. The Training Set consisted of \num{4500} apps, with 90 apps representing each of the 50 classes, with the Test Set comprising the remaining 500 apps. Since our \datasetName consists solely of benign apps, we expanded our Test Set by including 500 malicious apps sourced from the \az repository. We gathered apps that were flagged as malware by at least ten antivirus scanners in VirusTotal~\cite{virustotal}, and also those that were available on the Google Play Store, in order to easily access their descriptions.
    \item \textbf{Clustering.}
    Once we organized the dataset and retrieved the descriptions of all the apps, we proceeded to cluster the Training Set into 50 clusters using both the CHABADA original approach and \approachName. While clustering the apps, we made sure to save the machine learning models for future reuse during the testing process.
    \item \textbf{Sensitive APIs Extraction.}
    For all the apps in both the Training and Test sets, we extracted the sensitive API calls, which are API calls that require Android permission settings for protection and can access sensitive information (e.g., camera, microphone, etc.) or perform sensitive tasks (e.g., altering system settings, sending messages, etc.). For example, the \texttt{getLastKnownLocation()} API call is protected by \texttt{ACCESS\_FINE\_LOCATION} permission. To extract all sensitive API calls, including the number of call sites for each API, we used static analysis tools such as Androguard~\cite{androguard} and ApkTool~\cite{apktool}.
    \item \textbf{Training.}
    CHABADA leverages sensitive APIs as binary features and employs a One-Class SVM (OC-SVM) to train distinct models for each cluster of similar applications. These models are specifically tailored for detecting anomalies or novelties, which, in our case, refer to applications whose API usage significantly deviates from the established norm within their respective clusters. As described in the CHABADA paper, we trained an OC-SVM for each cluster of benign apps found in the training set, aiming to learn their "normal" behavior. We repeated this step first by using the clusters produced using the original CHABADA approach and then by using the clusters produced by \approachName (with a total of 50+50 models).  
    \item \textbf{Testing.}
    Finally, we used the cluster-specific OC-SVM models as malware detectors. We assign each app in the Test Set to one of the 50 clusters using the models saved from the clustering phase. This enables us to select the appropriate OC-SVM model to be used as an anomaly detector. Then, we verify whether the malicious apps are identified as anomalies, indicating differences in their APIs compared to the typical usage of the API within the same cluster. On the contrary, we expect the benign apps in the Test Set not to be flagged as anomalies i.e., potentially malicious, since their behavior should align with the common behavior of the specific cluster.
\end{enumerate}

\subsection{RQ5 Results.}
\label{subsec:RQ5}

\begin{table}[t]
\caption{Performance of CHABADA as a malware detector: original approach vs. \approachName ($F1 = \frac{TP}{TP + \frac{1}{2}(FP + FN)}$).}
    \begin{adjustbox}{width=.9\columnwidth,center}
\begin{tabular}{l|ll|ll|l}
             & \multicolumn{1}{c}{\textbf{TN Rate}} & \multicolumn{1}{c|}{\textbf{FP Rate}} & \multicolumn{1}{c}{\textbf{FN Rate}} & \multicolumn{1}{c|}{\textbf{TP Rate}} & \multicolumn{1}{c}{\textbf{F1}} \\ \hline
Chabada      & 86.40\%                              & 13.60\%                                & 80.40\%                              & 19.60\%                               & 0.29                                    \\
\approachName & \textbf{92.00\%}                     & \textbf{8.00\%}                       & \textbf{69.20\%}                     & \textbf{30.80\%}                      & \textbf{0.44}                          
\end{tabular}
    \end{adjustbox}
\label{table:RQ5}
\end{table}

The primary goal of RQ5 is to explore the potential advantages of \approachName when applied to tools reliant on automated app categorization, such as CHABADA. The results of our experiments are presented in Table \ref{table:RQ5}. As in the original CHABADA paper, we consider a malicious app detected as an anomaly as a True Positive, and a benign app not flagged as an anomaly as a True Negative~\cite{gorla-chabadaicse-2014}. The remaining definition of False Positive and False Negative, come easily. \approachName demonstrates improvements in both the False Positive Rate and the True Positive Rate. In the case of the False Positive Rate, we observed a slight reduction from 13.60\% to 8.00\%, indicating fewer benign apps being incorrectly identified as malicious. On the other hand, there has been a significant improvement in the True Positive Rate, which increased from 19.60\% to 30.80\%. This increase of 57.14\% clearly demonstrates the impact of \approachName in identifying a larger number of malicious apps, when CHABADA relies on it. In terms of overall performance, the F1 Score has increased from 0.29 to 0.44, confirming the positive impact of \approachName, as just mentioned. Moreover, it is important to note that the OC-SVM models, which do not have any prior knowledge about malicious apps, can also be used to detect unknown malware, making the results even more remarkable.

\begin{formal}
\textbf{Answer to RQ5:} \textit{We demonstrated that relying on a better categorization approach, such as \approachName, can yield a substantial influence on the final outcome of tools that depend on categorization, such as CHABADA.}
\end{formal}

%% file: Sections/08_Limitations.tex
\section{Limitations and Threats to Validity.}
Like any other study, our research is susceptible to threats to validity, which arise from various limitations in our approach. Below, we outline the most significant threats and limitations.

\noindent\textbf{Human error and subjectivity.} 
Constructing the first ground-truth dataset from scratch posed challenges as we could only rely on manual verification to ensure high-quality data. It is important to note that despite following a consistent process for building \datasetName, human subjectivity may still influence certain aspects. To mitigate this threat to validity we share \datasetName, as well as the tools and scripts used to create it, with the research community.

\noindent\textbf{Not exhaustive literature search.} 
Our literature search for categorization approaches, based on the six queries we defined in ~\ref{subsec:literatureSearch}, may not be exhaustive. However, as it serves as a systematic means to identify existing approaches in a structured manner, our decision was primarily driven by practical considerations. Indeed, including broader queries would have been impractical due to the overwhelming number of papers retrieved. For instance, the query "app clustering" returns over \num{200000} results on Google Scholar, making it impractical to manually select the relevant approaches for evaluation.

\noindent\textbf{Approaches selection and code availability.}
The selection of categorization approaches for evaluation, carried out in Section~\ref{subsec:approachSelection}, was mainly influenced by code availability. The lack of code hindered a comprehensive performance overview, highlighting how much important it is to provide code alongside the paper to benefit the entire research community.


%% file: Sections/09_RelatedWork.tex
\section{Related Work}

To the best of our knowledge, we are the first to provide a comprehensive ground-truth dataset for evaluating Android app categorization. However, we now present a concise summary of the primary findings derived from previous research conducted on this topic.

The previous research conducted by Al-Subaihin et al.~\cite{alsubaihin2019} is highly relevant to our work as it presents an empirical comparison of various text-based app clustering techniques such as Latent Dirichlet Allocation (LDA), as well as keyword feature extraction methods. They experimented with \num{12664} Google Play Store apps sampled from 24 categories. They utilized intrinsic measures such as the Silhouette Score and human judgment (analyzing 300 apps in total) due to the absence of a  ground-truth, as outlined in their paper. Our work assumes even greater significance due to the creation of \datasetName, which could have proven immensely valuable in evaluating studies like the one conducted by Al-Subaihin et al.~\cite{alsubaihin2019}. Furthermore, in our paper, we not only assessed approaches that rely solely on app descriptions for categorization but also tried to distinguish between the features utilized by the evaluated methods. For instance, we took into consideration all types of data that can be extracted from the APK in order to provide a comprehensive evaluation.

Martin et al.~\cite{martin2017} conducted a comprehensive survey on App Store Analysis, which included a dedicated section on App Clustering approaches. In this section, they presented an overview of various techniques and features utilized in clustering apps highlighting how the app descriptions are commonly leveraged to categorize apps according to their functionality. In their review, about Android Security using NLP, Sen et al.~\cite{sen2021} provide a concise overview of categorization approaches commonly employed in malware detection tasks, which typically rely on the descriptions of the apps. Although our literature search may not have been as comprehensive as the studies conducted by Martin et al.~\cite{martin2017} and Sen et al.~~\cite{sen2021}, we have gone beyond by not only offering an overview of existing approaches but also assessing their performance on \datasetName ground-truth.

Several research papers have consistently highlighted the inadequacy of Google Play's current app categorization system. Apps within the same category often exhibit only a vague sense of similarity, indicating a significant lack of effective categorization~\cite{alsubaihin2016, martin2017, gorla-chabadaicse-2014, surian2017}. In our paper, we conducted a comprehensive analysis of our own original dataset affirming the findings of the just cited prior studies. We even provided concrete evidence of a misclassified app, as defined by Surian et al. in their paper~\cite{surian2017}.

%% file: Sections/10_Conclusion.tex
\section{Conclusion}
\label{sec:conclusion}

In our study, we conducted a comprehensive evaluation of various Android app categorization approaches found in the existing literature. Our analysis emphasized the remarkable superiority of approaches that utilize app descriptions, as opposed to those relying solely on data extracted from the APK file. This evaluation was made possible thanks to our new \datasetName ground-truth. We believe this dataset will be a valuable resource for future research in this field, addressing a gap that has been previously highlighted by similar studies ~\cite{alsubaihin2019, ebrahimi2021}.
Furthermore, we developed two innovative approaches that effectively improve the performance of existing methods in both description-based and APK-based methodologies. Notably, for what concerns our description-based approach \approachName, we achieved an impressive ARI score of \bestAriScoreDescription by leveraging the powerful text embedding models provided by OpenAI. In our final experiments, we demonstrated the impact of a better-performing categorization approach when implemented within a tool reliant on automatic app categorization, such as CHABADA~\cite{gorla-chabadaicse-2014}. This highlights how \approachName can provide substantial benefits to future software engineering tools that rely on automatic categorization, emphasizing the importance of developing advanced and efficient app categorization methodologies.

Future work may involve expanding \datasetName by incorporating additional apps and classes, as well as intensifying efforts to address the existing disparity between APK-based and description-based approaches. By offering valuable insights, introducing a new ground-truth dataset, and presenting two innovative approaches to address existing limitations in the literature, our research aims to make a significant contribution to the field of automatic app categorization in software engineering.

%% file: Sections/11_DataAvailability.tex
\section{Data Availability}
\label{sec:data_availability}

The repository including all artifacts is available at: 
\begin{center}
   \href{https://github.com/Trustworthy-Software/Revisiting-Android-App-Categorization}{https://github.com/Trustworthy-Software/Revisiting-Android-App-Categorization}
\end{center}

%% file: Sections/12_Acknowledgement.tex
\section{Acknowledgement}
\label{sec:Acknowledgement}
This research was funded in part by the Luxembourg National Research Fund (FNR), grant reference NCER22/IS/16570468/NCER-FT and REPROCESS grant reference C21/IS/16344458.